\documentstyle[prl,aps,epsfig,multicol]{revtex}
\newcommand{\be}{\begin{equation}}
\newcommand{\ee}{\end{equation}}

\newcommand{\bea}{\begin{eqnarray}}
\newcommand{\eea}{\end{eqnarray}}
\newcommand{\MeV}{{\rm MeV}}
\newcommand{\eV}{{\rm eV}}
\newcommand{\srm}{{\rm s}}
\newcommand{\Be}{{\rm Be}}
\newcommand{\Bo}{{\rm B}}
\newcommand{\Cl}{{\rm Cl}}
\newcommand{\Ga}{{\rm Ga}}
\newcommand{\SK}{{\rm SK}}
\newcommand{\SNO}{{\rm SNO}}
\newcommand{\CC}{{\rm CC}}
\newcommand{\SNOCC}{{\rm SNO-CC}}
\newcommand{\expt}{{\rm expt}}
\begin{document}
\title{Constraints on  decay plus oscillation solutions of the solar
neutrino problem }
\author{
Anjan S. Joshipura$^a$, Eduard Mass{\'o}$^b$ and  Subhendra Mohanty$^a$}

\address{ $^a$Physical Research Laboratory,
Navrangpura, Ahmedabad - 380 009, India\\
$^b$Grup de F{\'\i}sica Te{\`o}rica and IFAE,
Universitat Aut{\`o}noma de Barcelona,
 Spain.}

%\date{\today}
%\date{}

\maketitle

\begin{abstract}
We examine the constraints on non-radiative decay of neutrinos
from the observations of solar neutrino experiments. The standard
oscillation hypothesis among three neutrinos solves the solar and
atmospheric neutrino problems. Decay of a massive neutrino mixed
with the electron neutrino results in the depletion of the solar
neutrino flux. We introduce neutrino decay in the oscillation
hypothesis and demand that decay does not spoil the successful
explanation of solar and atmospheric observations. We obtain a
lower bound on the ratio of the lifetime over the mass of $\nu_2$,
$(\tau_2/m_2) > 22.7\, (\srm/\MeV)$ for the MSW solution of the
solar neutrino problem
 and  $(\tau_2/m_2) > 27.8\, (\srm/\MeV)$  for the VO solution (at
$99\%$ C.L.).

\end{abstract}

%\begin{multicols}{2}

\section{Introduction}

 Solar neutrino experiments with Chlorine \cite{cl}, Gallium
\cite{ga}, and water cerenkov detectors  \cite{sk-sol},\cite{sno},
show unequivocally that there is a deficit of $\nu_e$ at the Earth
compared to the predictions of the standard solar model
\cite{bp2000}. It is commonly accepted that vacuum oscillations
(VO) or matter induced MSW conversions, with large mixing angle
(LMA) or small mixing angle (SMA), can account for the deficit of
solar neutrinos. The available experimental results allow not only
a test of the neutrino oscillation hypothesis but also offer
possibilities of constraining new physics, {\it e.g.} neutrino
magnetic moment \cite{moment}, neutrino decay, flavour changing
neutral currents, etc. Here we will be concerned with neutrino
decay in the context of neutrino oscillations. The possibility of
solving the solar neutrino problem only through neutrino decay in
vacuum was raised in \cite{ack} and ruled out mainly owing to the
fact that the lower energy $pp$ neutrinos observed in Gallium
experiments are less suppressed compared to higher energy $^7$Be
and $^8$B neutrinos observed in Chlorine experiments. The
suppression in neutrino flux caused by the solar matter induced
decay to majoron has correct energy dependence \cite{zurab} but
the required fast rates cannot easily be obtained in the standard
scenario without conflicting with other constraints \cite{zurab}.
The oscillation plus vacuum decay scenario has been studied more
recently in a two generation model in \cite{sg-vo},\cite{sg-msw}.
This analysis was prior to results from SNO \cite{sno}. These
results in combination with earlier data from SuperKamioka
\cite{sk-sol} can be used to separate the flux of the electron
neutrino  from that of other active flavours. This additional
information provided by SNO can be quite useful in probing new
physics. In the light of this, we study in this paper the three
generation model of neutrino oscillation plus decay including  the
recent SNO \cite{sno} result in our analysis.

Radiative decays of neutrinos are severely constrained by
laboratory experiments and a variety of astrophysical and
cosmological observations (see references in the PDG \cite{PDG}).
However, constraints on non-radiative decays are much less
stringent. For a relatively heavy unstable neutrino, one can apply
the argument that decay products would contribute to the energy
density of the universe and thus obtain a limit \cite{mp}.
However, for masses of the order of  eV or less, there is no limit
on the lifetime. Solar neutrino data constrains non-radiative
neutrino decay in this mass regime.

We examine the scenario where the neutrinos from the Sun are
depleted due to non-radiative decays like (a) Majoron emission
decays $\nu_2 \rightarrow \bar \nu_1 +J$, where the $\bar \nu_1$
state is either sterile or the active antiparticle of $\nu_1$, or
(b) $\nu_2 \rightarrow 3 \nu$. We assume that the lowest mass
neutrino $\nu_1$ is stable, or at least that it has lifetime
$(\tau_1/m_1) \gg 20\, (\srm/\MeV)$. (In the paper, we denote by
$m_i$ and $\tau_i$ the mass and rest frame lifetime of the $\nu_i$
neutrino.) The neutrinos which arise as decay products in (a) or
(b) can be either active or sterile. They are depleted in energy
and carry on an average less than half of the solar neutrino
energy. Since the flux of the electron neutrino sharply increases
as we go down in energy, the relative contribution of the
neutrinos produced in decay to the total signal in a given
experiment will be quite small even if they are active. Because of
this reason the net effect of the decay is depletion in the
neutrino flux. This depletion is over and above the one caused by
the neutrino oscillations or conversion which we assume to be the
main cause of the solar deficit. The  additional depletion caused
by the decay is given by an exponential function \be \exp\left(-
{R \over \tau_2}{m_2\over E_\nu}\right) \ee where $E_\nu$ refers
to the energy of the decaying neutrino and $R$ is the Sun-Earth
distance. The solar neutrino survival probabilities are given in
terms of the mixing $s_1$, mass difference $\delta m^2 \equiv
m_2^2-m_1^2$ and the decay lifetime $\tau_2$. We determine the
regions in the $\tan^2\theta_1-\delta m^2$ plane allowed at $99\%$
C.L. for different values of $\tau_2$. The allowed region is found
to shrink as $\tau_2$ decreases and ultimately it disappears at
some value which is taken to be the bound on the neutrino lifetime
to invisible channel. The bound derived this way depends upon the
specific solution of the solar neutrino problem and is stronger in
case of the VO solution $(\tau_2 /m_2) > 27.8\, (\srm/\MeV)$
compared to the MSW-LMA and SMA solutions for which we get
$(\tau_2 /m_2) > 22.7\, (\srm/\MeV)$. In an earlier pre-SNO
analysis \cite{sg-msw},\cite{sg-vo}, similar bounds were obtained
for the LMA and vacuum solutions but there was no bound on
$\tau_2$  for the SMA solution. We show in the final section that
inclusion of SNO result is critical for obtaining a bound on
$\tau_2$.

We assume the mixing among three neutrinos to be responsible for
the deficit in the neutrino flux observed by the solar and
atmospheric neutrino experiments. The mixing matrix $U$ needed in
order to accomplish this has the following form:
 \bea
\pmatrix{
  \nu_e \cr
  \nu_\mu\cr
  \nu_\tau}
= \pmatrix{
  c_1& s_1 & 0 \cr
  -c_2 s_1 & c_2 c_1 & s_2\cr
  s_2 s_1& -s_2 c_1& c_2}
 ~~\pmatrix{
  \nu_1 \cr
  \nu_2\cr
  \nu_3}
 \label{u}
\eea
We have approximated $U_{e3} \sim 0$ in order to account for the
negative results obtained in  $\nu_e$ disappearance experiment at
CHOOZ \cite{chooz}. In addition to the above mixing one also needs
$\delta m^2= m_2^2-m_1^2\approx 10^{-4}-10^{-11}$
eV$^2$ and $\delta_A\equiv m_3^2-m_2^2\approx 10^{-3}$
eV$^2$ to account for the solar and atmospheric
\cite{sk-atm} neutrino scales.
The required value of $\delta_A$ is much larger than the
value of the effective mass square $2\sqrt{2}G_FEN_e$ of the
electron
neutrino at the solar core. This suppresses mixing of the third
neutrino in matter \cite{kras} which decouples from the rest in
case of the MSW solution to the solar neutrino problem. The small
value of $U_{e3}$ (taken here as zero) also prevents the mixing of
the third neutrino in vacuum. As a result, the solar data cannot
be used to constrain the lifetime of the heaviest mass eigenstate
lying at the atmospheric scale. We therefore concentrate on
limiting the $\nu_2$ lifetime.

\section{VO Probabilities}

Neutrinos are produced and detected as flavor eigenstates
$\nu_\alpha , \alpha =e,\mu,\tau$ but their time evolution
operator is diagonal in the mass basis $\nu_i, i=1,2,3$. The
amplitude for flavor conversion during vacuum propagation is given
by
\be
{\cal A}_{\alpha \beta}(t) = {\displaystyle \sum_{i}} U_{\alpha
i}U^*_{i \beta} {\cal A}_i(t) \ee
 where $U_{\alpha i}$ are elements of the mixing angle matrix and
 ${\cal A}_i(t)$ is the time evolution operator for the $\nu_i$
 mass eigenstate. For the decaying neutrino scenario the time
 evolution amplitude is given by
 \be
 {\cal A}_i(t) = \exp(-E_i t)\
\exp\left( -{t \over 2 \tau_i}{m_i\over E_i}\right)
 \ee
 where we allow for decay of the $\nu_i$ mass
 eigenstate with lifetime $\tau_i$
and we approximate $E_i \simeq p +m_i^2/(2p)$. The probability for
 flavor conversion during propagation in vacuum at time $t$ is
 then given by
 \be
 P_{\alpha \beta}(t) ={\displaystyle \sum_{i>j}} U_{\alpha i}
 U^*_{i \beta} U_{j \beta} U^*_{\alpha j} \
\cos\left[{(m_i^2 -m_j^2)  t\over 2 E_\nu} \right]\
\exp\left[ -\left( {m_i\over 2 \tau_i} +{m_j \over 2 \tau_j} \right)
{t\over E_\nu} \right]
\ee
with $E_\nu=p$.

Now we
assume that the lightest mass state $\nu_1$
does not decay.
Using the mixing matrix (\ref{u}) the $\nu_e \rightarrow \nu_e,
\nu_\mu, \nu_\tau$ conversion probability in vacuum from Sun to
Earth is given by the expressions \bea
 P_{ee} &=& c_1^4 + s_1^4~ \exp\left(-{ \alpha
 \over E_{\nu}}\right) + 2 (c_1 s_1)^2~ \cos\left({\delta m^2 R \over 2
 E_{\nu}}\right)~ \exp\left(-{ \alpha
 \over 2 E_{\nu}}\right) \nonumber\\
 P_{e \mu}&=& c_1^2 s_1^2 c_2^2\ \left[1 + ~\exp\left(-{ \alpha
 \over E_{\nu}}\right) -2
~\cos\left({\delta m^2 R\over 2
 E_{\nu}}\right)~ \exp\left(-{ \alpha
 \over 2 E_{\nu}}\right) \right]\nonumber\\
P_{e \tau}&=&c_1^2 s_1^2 s_2^2\  \left[1 + ~\exp\left(-{ \alpha
 \over E_{\nu}}\right) -2
~\cos\left({\delta m^2 R\over 2
 E_{\nu}}\right)~ \exp\left(-{ \alpha
 \over 2 E_{\nu}}\right)\right]
 \label{pvo}
 \eea
where \be \alpha= { R\,  m_2 \over \tau_2 } \ee $R=1.5 \times
10^{13}\, {\rm cm}\, =500.3$ s being the Earth-Sun distance.
We notice that the
$\nu_3$ decay lifetime does not occur in the expressions
(\ref{pvo}) as $U_{e 3} \simeq 0$. The
charged and the neutral current interactions in the detector
respectively are related to $P_{ee}$ and $P_{e\mu}+P_{e\tau}$. Both of
these are seen to be independent of the atmospheric mixing angle.
The sum of all three probabilities is also independent of the solar
$\delta m^2$,
\be
 P\equiv P_{ee}+P_{e \mu}+P_{e \tau}= c_1^2 + s_1^2
\exp\left(-{ \alpha \over E_{\nu}}\right)
 \label{sum1}
 \ee

\section{MSW probabilities}

If the flavor conversion of neutrinos in the Sun is by MSW
mechanism then we have  different expressions for probabilities
for the decay plus conversion scenario. As discussed already, only
two energy eigenstates $\nu_1$ and $\nu_2$ corresponding to the
lighter neutrinos participate in the MSW conversion. In the core
of the Sun $E_1
> E_2$, there is a level crossing at the resonance point after
which $E_2 > E_1$ {\it i.e.} $m_2 > m_1 $ in vacuum.

The probability of $\nu_e \rightarrow \nu_1$ just after level
crossing is
\be
P_1  =P_J s_m^2 + (1-P_J) c_m^2
\ee
Here, the first term stands for $\nu_e$ going to $\nu_2$ by  mixing with
probability $s_m^2$ and then jumping to $\nu_1$ with probability
$P_J$ at the level crossing. The second term means that $\nu_e$
goes to $\nu_1$ by mixing and then does not jump  to $\nu_2$ with
probability $(1-P_J)$ at the level crossing.
The probability of $\nu_e \rightarrow \nu_2$ just after
level crossing is
\be
P_2 =(1-P_1)=(1-P_J) s_m^2 + (P_J) c_m^2 \ee

In the formulas above, the Landau-Zener jump probability is given
by

\bea P_{J} &=& {\exp(-b s_1^2 /E_\nu)-\exp(-b /E_\nu)\over
1-\exp(-b /E_\nu)} \label{lz}\\ b&=&  {\pi \over 4}\,
\left({\delta m^2   \over
 |\dot A /A|_{res}} \right)
\simeq 10^9\, \left({\delta m^2 \over \eV^2}\right)\, \MeV
\eea
 and $A= 2 \sqrt{2} E_\nu G_F N_e$.
The mixing angle in matter in the Sun is given in terms of the
vacuum mixing angle by the expression \bea \cos 2\theta_m &=&
{(-1+\eta(1-2 s_1^2))\over (1- 2 \eta(1-2 s_1^2) +\eta^2)^{1/2}} \\
\eta&=& { \delta m^2 \over A} =6.6\times 10^{-5} {b\over E_\nu}
\eea

After level crossing the $\nu_1$ state stays as it is but the
$\nu_2$ state can decay into antineutrinos or sterile neutrinos.
Thus,
at Earth the probability of detecting $\nu_1$ is $P_1$ and  of detecting
$\nu_2$ is $P_2\exp(-\alpha/E)$.
We can now use the matrix (1) to find the $\nu_e , \nu_\mu, \nu_\tau$
content of neutrinos at Earth,
\bea
P_{ee}&=& c_1^2 P_1 + s_1^2 P_2 \exp(-\alpha /E_\nu) \nonumber \\
P_{e \mu}
&=& c_2^2 s_1^2 P_1 + c_2^2 c_1^2 P_2 \exp(-\alpha /E_\nu)
\nonumber \\ P_{e
\tau}&=&s_2^2 s_1^2 P_1 + s_2^2 c_1^2 P_2 \exp(-\alpha /E_\nu)
\label{pmsw} \eea As in case of the VO, both $P_{ee}$ and
$P_{e\mu}+P_{e\tau}$ are independent of the atmospheric angle. But
unlike VO, the  sum of the three probabilities in (\ref{pmsw}) \be
  P\equiv P_{ee}+P_{e \mu}+P_{e \tau}=P_1 +(1-P_1)
\exp(-\alpha/ E_{\nu})
   \label{sum2}
 \ee
depends on the solar scale through the Landau-Zener formula
(\ref{lz}).

\section{Experimental rates and bound on lifetime}

We use the probabilities derived above to obtain a bound
on the neutrino
lifetime. We consider only total rates for this purpose and
include all experiments in our analysis. We do the analysis by two
different methods. First we consider the total rates in Chlorine \cite{cl},
Gallium \cite{ga} , Super-K \cite{sk-sol} and SNO \cite{sno}
experiments and determine the allowed regions in the $tan^2 \theta_1
 -\delta m^2$ parameter space for different values of $\tau_2$.
The $Cl$, $Ga$ and the charged current rates of SNO by themselves
cannot constrain $\tau_2$ in the SMA region of the MSW solution
since these experiments measure $P_{ee}$ which becomes independent
of $\tau_2$ in the small mixing angle limit. This is not true
however in case of $P_{e \mu} +P_{e \tau}$ which is probed by
neutral current events in SK. Since the neutral current rates
can be inferred by combining SK and SNO data , such combination is
expected to constrain the lifetime $\tau_2$. We show this
explicitly by using the SK and SNO data alone.

The rates of neutrino capture in the Chlorine and Gallium
experiments can be written as \be
 R_\alpha ={ {\displaystyle \sum_{i=pp,\Be,\Bo}}
\int dE_\nu \Phi_i  \sigma_{ \alpha} P_{ee}\over
{\displaystyle \sum_{i=pp,\Be,\Bo}} \int dE_\nu \Phi_i  \sigma_{ \alpha}}
 \ee
where the subscript $\alpha=$ Ga, Cl denotes the experiment and
$i=pp$, $^7$Be, $^8$B denotes the type of neutrino flux from the Sun.
The spectra of the $pp $ and $^8$B neutrinos can be fitted with
the analytical functions,
\bea
\Phi_{pp} &=&(5.95\times 10^{10})
[193.9 (0.931-E_\nu)((0.931-E_\nu)^2 -0.261)^{1/2} E_\nu^2]
\nonumber \\
 \Phi_{\Bo}&=&(5.05 \times 10^{6} )[8.52
\times 10^{-6}(15.1-E_\nu)^{2.75} E_\nu^2 ]  \nonumber
\\
\Phi_{\Be} &=&(4.77 \times 10^{9}) [ \delta(E_\nu -0.862)]
\label{flux}
 \eea
where the neutrino fluxes are in units of cm$^{-2}$s$^{-1}$ and
$E_\nu$ is
in MeV. The first brackets in (\ref{flux}) give the
total flux of neutrinos from the $pp$,Be, and B reactions
and are taken from
BP2000 \cite{bp2000}, and the square brackets give the  spectral
shape \cite{bu}.

The Ga experiments can detect all three types of neutrino fluxes and the
neutrino absorbtion cross section of $Ga$ is given in
\cite{Ba-Ga}. The Chlorine experiment threshold is higher (0.8 MeV) and it
detects only the Be and B neutrinos; we take the absorption cross
section with the tables from ref.\cite{Ba-Cl}.

 The electron scattering reaction in Super-K and
the charge-current deuterium dissociation  reaction at SNO can be
written as
\bea
R_{\SK} = {\int dE_\nu \sigma_{\nu_e} \Phi_\Bo P_{ee} +\int
dE_\nu \sigma_{\nu_\mu} \Phi_\Bo (P-P_{ee})\over \int dE_\nu
\sigma_{\nu_e} \Phi_\Bo } \label{tsk}
\eea
 and
\bea
R_{\SNO}^{\CC}= {\int dE_\nu \sigma_{\CC} \Phi_\Bo P_{ee} \over
\int dE_\nu \sigma_{\CC} \Phi_\Bo } \label{tsno}
\eea
The $\nu_e e^-$ and $\nu_{\mu,\tau} e^-$ elastic scattering cross
section after folding with the detector response function are
tabulated in ref.\cite{Ba-SK}. The deuterium dissociation cross section
are taken from \cite{Ba-snocc}.

Using the flux spectrum in equation (\ref{flux}) and the cross
sections \cite{Ba-Ga,Ba-Cl,Ba-SK,Ba-snocc}
we can calculate the
theoretical rates for MSW or VO conversion probabilities as a
function of the three unknown parameters: the $\nu_2$ lifetime
$\tau_2$,
$\delta m^2$ and the vacuum mixing angle $\theta_1$.
The experimental rates $R_\alpha^{\expt}$ with one-sigma combined
(statistical and systematic) experimental errors $\Delta_\alpha$
are as follows \cite{cl,ga,sk-sol,sno}:
 \bea
 R_{\Cl}^\expt &=& 0.335 \pm 0.029 \nonumber\\
 R_{\Ga}^\expt &=& 0.584 \pm 0.039 \nonumber\\
 R_{\SK}^\expt &=& 0.459 \pm 0.017 \nonumber\\
 R_{\SNOCC}^\expt &=&0.347 \pm 0.029
 \label{rexp}
\eea
From the theoretical $R_\alpha$, and the experimental
$R_\alpha^{\expt}$,
we compute the total $\chi^2$ for all experiments,
defined as
\be
\chi^2 \equiv {\displaystyle \sum_{\alpha=
\Cl,\Ga,\SK,\SNO}}{(R_\alpha -R_\alpha^{\expt} )^2 \over
\Delta_\alpha^2} \ee.

Setting the parameter $\alpha =0$ we reproduce the standard
contours of LMA , SMA and VO solutions shown as dotted contours in
Fig. 1 and Fig. 2. For $\alpha=0$ the global minimum of  $\chi^2$
occurs in the vacuum region with  $\chi_{min}^2\sim 0.3$.The
$\chi^2_{min} +11$ contours which corresponds to 99\% C.L. bounds
for $\alpha$ are plotted for various values of non-zero $\alpha$
as shown in the solid contours in Fig.~1 and Fig.~2. We find that
the LMA and SMA allowed parameter space disappear for $\alpha =18$ and
$\alpha=22$ (in MeV units)respectively. The VO allowed region disappears
for
$\alpha= 18$. These numbers translate into the bounds  $\tau_2 > 22.7\,
\srm\, (m_2/\MeV)$ for the MSW solution (where we have taken the larger 
of the two bounds coming from SMA and LMA solutions) and $\tau_2 > 27.8 \,
\srm\, (m_2/\MeV)$ for the VO solution.

As discussed  before the bound on $\tau_2$ in the SMA region
mainly comes from the SK data which probe $P_{e \mu} +P_{e \tau}$.
It is thus interesting to consider only the SK and SNO results
which together determine $P_{ee}$ and $P_{e \mu}+P_{e \tau}$ in
the same energy range ($8-15$ MeV). These two experiments are
crucial in providing the bound on $\tau_2$. To see the impact of
these results , we plot the $1.96 \sigma$ contours of the SNO
($R_{\SNO}=0.347 \pm 1.96*0.029$ shown as dashed curves) and SK
($R_{\SK}= 0.459 \pm 1.96*0.017$ shown as continuous curves)
rates. In Fig.3 we show the allowed regions for $\alpha=0$. In
Fig.4 we plot the same SNO and SK rates for $\alpha=25 $ and we
can see that there is no overlap in the SMA region between the two
experiments for this value of $\alpha$. The bound on $\alpha$
obtained from SK and SNO is  marginally weaker than the one
obtained in the combined analysis of all  (SK,SNO, Ga and Cl) experiments.

In summary, we have used available results of the solar and
atmospheric neutrino experiments to constrain neutrino lifetime
within the three generation picture of neutrino oscillations.
While the exact bound depends upon the specific solution, we
typically find ${\tau_2\over m_2}> 28  \left({sec\over MeV}\right)
$ The corresponding bound $\tau_2>2.8 \cdot 10^{-5} sec$ for eV mass
neutrinos is  not very strong but useful since it is the only one
following from laboratory experiments in this  mass range.

 {\it Acknowledgements}: The work of E.~M. is partially supported
by the CICYT Research Project AEN99-0766, by the DGR Project 2001
SGR 00188, and by the EU network on {\it Supersymmetry and the
Early Universe} (HPRN-CT-2000-00152).

%\end{multicols}

\newpage

\begin{figure}
 \label{Fig.1}
\centering
\includegraphics[width=10cm]{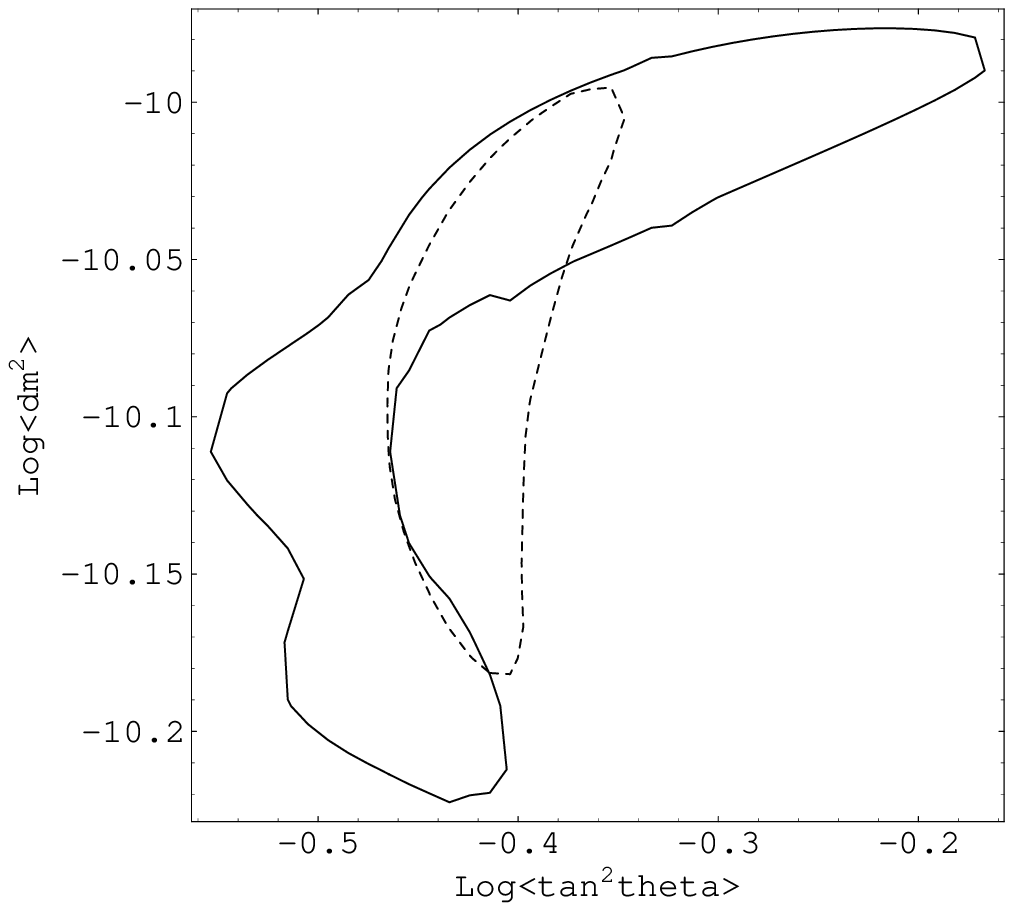}
\caption{ Allowed parameter space for VO plus decay solution, continuous
curve is for decay
parameter $\alpha=0$
and dashed curve is for $\alpha=10$.}
\end{figure}

\begin{figure}
 \label{Fig.2}
\begin{center}
\includegraphics[width=10cm]{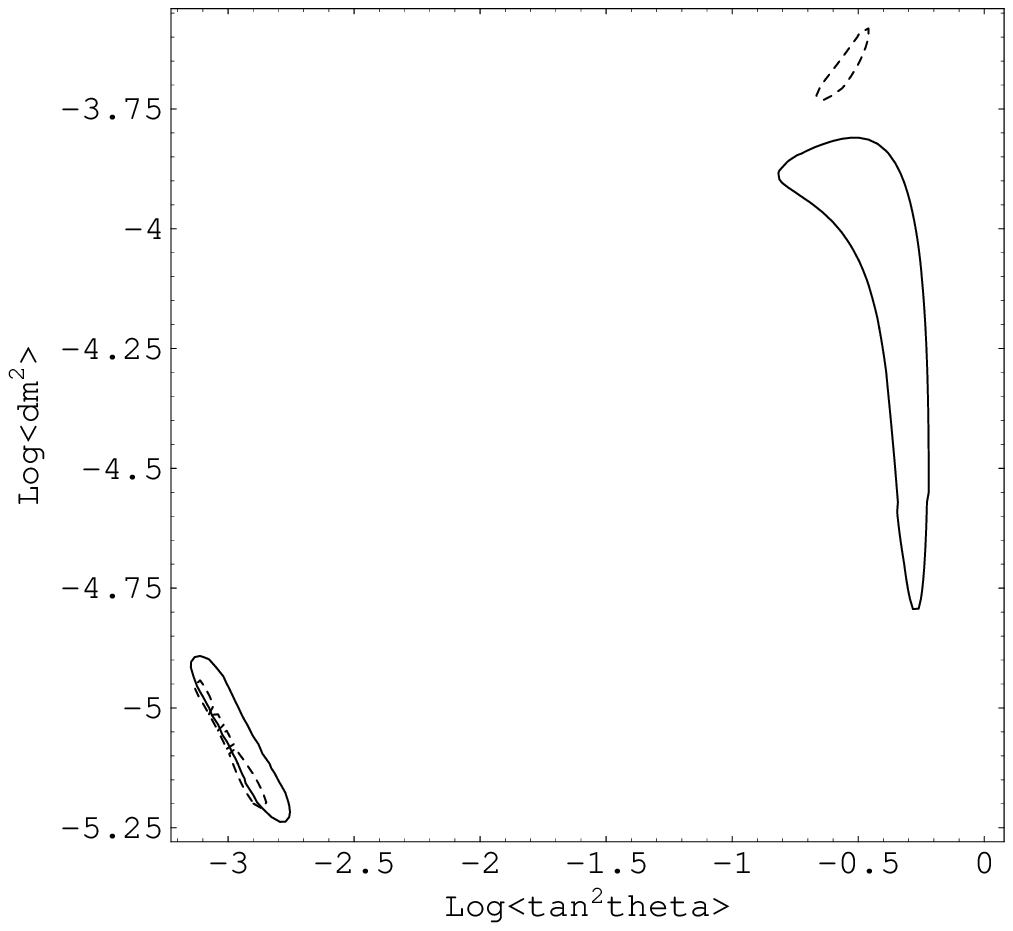}
\end{center}
\caption{  Allowed parameter space for the MSW plus decay solution,
continuous curve is for decay
parameter $\alpha=0$
and dashed curve is for $\alpha=10$.  }
\end{figure}

\newpage

\begin{figure}
 \label{Fig.3}
\begin{center}
\includegraphics[width=10cm]{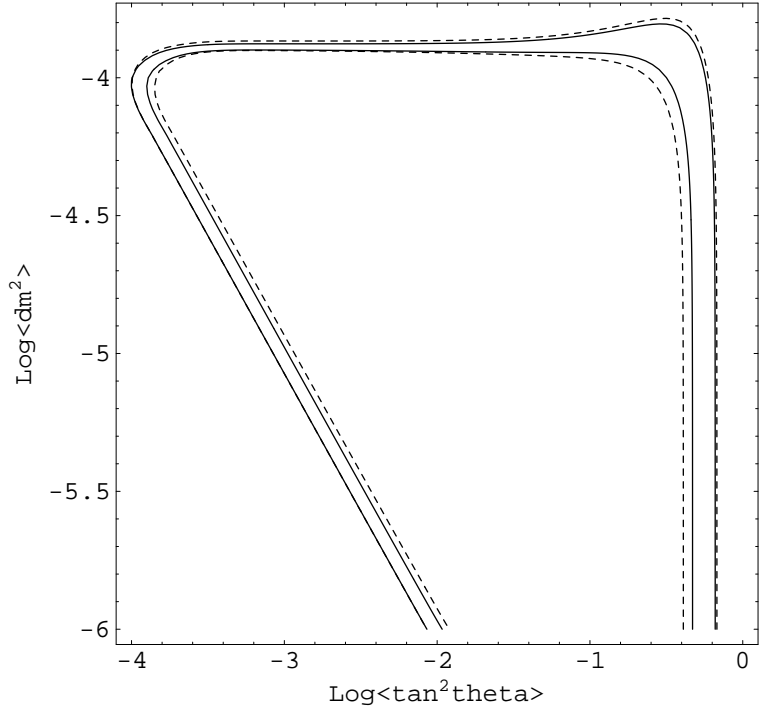}
\end{center}
\caption{The $1.96 \sigma$ allowed parameter space for
$R_{SK}$ as region enclosed by  dashed curves and $R_{SNO}$ as region
enclosed
by continuous curves; for decay parameter $\alpha=0$. } \end{figure}

\begin{figure}
 \label{Fig.4}
\begin{center}
\includegraphics[width=10cm]{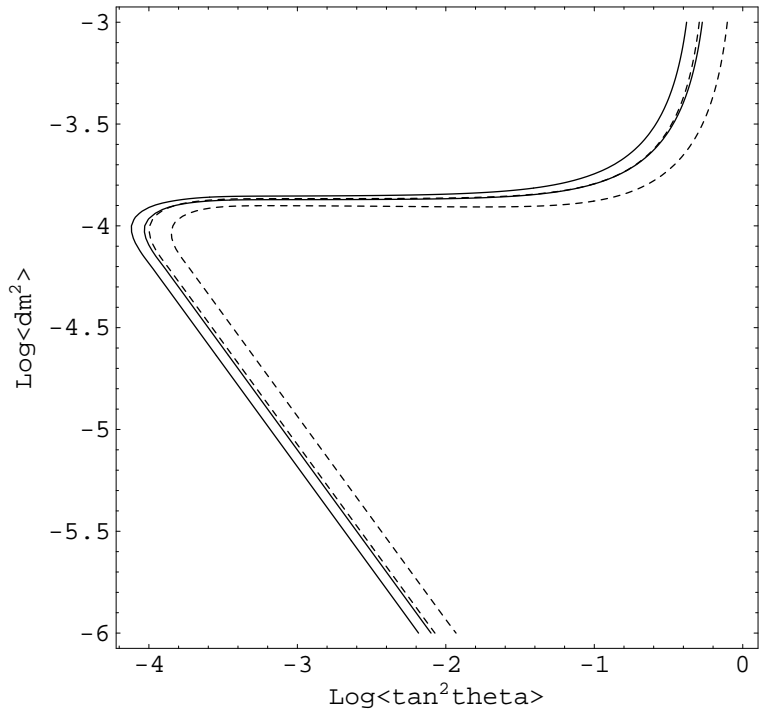}
\end{center}
\caption{The $1.96 \sigma$ allowed parameter space for
$R_{SK}$ as region enclosed by  dashed curves and $R_{SNO}$ as region
enclosed
by continuous curves; for decay parameter $\alpha=25$. There is
no overalap
in the SMA region for $\alpha=25$.}
\end{figure}

\end{document}